\documentclass{esagnc}

\usepackage{placeins}
\usepackage{float} 

\newcommand{\sing}{\ensuremath{\text{sing}}}

\begin{document}

\papertitle{Investigation of the Robustness of Neural Density Fields}

\mainauthor{J. Schuhmacher}     
\author[(1)]{Jonas Schuhmacher}
\author[(1)]{Fabio Grat}
\author[(2)]{Dario Izzo}
\author[(2, 3)]{Pablo Gómez}

\affil[(1)]{ \textit{Technische Universität München, Arcisstraße 21, 80333 München, Germany, \{jonas.schuhmacher, f.gratl\}@tum.de}}
\affil[(2)]{\textit{Advanced Concepts Team, European Space Agency, European Space Research and Technology Centre (ESTEC), Keplerlaan 1, 2201 AZ Noordwijk, The Netherlands, \{pablo.gomez, dario.izzo\}@esa.int}}
\affil[(3)]{\textit{AI Sweden, Lindholmspiren 11, 417 56 Göteborg, Sweden}}

\paperabstract{
Recent advances in modeling density distributions, so-called neural density fields, can accurately describe the density distribution of celestial bodies without, e.g., requiring a shape model - properties of great advantage when designing trajectories close to these bodies. 
Previous work introduced this approach, but several open questions remained. This work investigates neural density fields and their relative errors in the context of robustness to external factors like noise or constraints during training, like the maximal available gravity signal strength due to a certain distance exemplified for 433 Eros and 67P/Churyumov-Gerasimenko.
It is found that both models trained on a polyhedral and mascon ground truth perform similarly, indicating that the ground truth is not the accuracy bottleneck. The impact of solar radiation pressure on a typical probe affects training neglectable, with the relative error being of the same magnitude as without noise. However, limiting the precision of measurement data by applying Gaussian noise hurts the obtainable precision. Further, pretraining is shown as practical in order to speed up network training. Hence, this work demonstrates that training neural networks for the gravity inversion problem is appropriate as long as the gravity signal is distinguishable from noise.

Code and results are available at \url{https://github.com/gomezzz/geodesyNets}
}


\section{Introduction}

In recent years small bodies in our solar system have been of increasing interest as mission targets. Various past, present, and future missions visited them and even collected samples allowing study of their composition and properties. For example, in the past, NEAR visited the asteroid 253 Mathilde and later soft landed on 433 Eros in February 2001, or Rosetta, with its lander Philae visiting the comet 67P/Churyumov–Gerasimenko in 2014.
There are also several upcoming missions such as Hera, ZhenHe, and Psyche \cite{izzo2021geodesy}.
Hera, the complementary mission of ESA to NASA's DART mission, will analyze the latter's impact with the smaller moon Dimorphos, which revolves around the asteroid Didymos, with the goal of studying the process of asteroid deflection. It is scheduled to start in 2024 and rendezvous the binary system in 2026 together with two accompanying cubesats \cite{trisolini2023target}.

With these prospects, it is even more crucial for guidance, navigation and control to have accurate and precise models of these target bodies. However, this is especially difficult with small bodies since of their irregular shape. To complicate matters, the density distribution of these bodies is also rarely homogeneous but rather heterogeneous in nature \cite{leonard2019osiris, scheeres2020heterogeneous, fujiwara2006rubble}.

These facts make modeling these bodies with the established three models difficult: spherical harmonics, mascon models and polyhedral gravity models. The former struggle with the convergence inside the Brillouin sphere, and the irregular shapes slow down convergence, making it unsuitable for asteroids and comets \cite{vsprlak2021use, izzo2021geodesy, furfaro2021modeling}. The minimum Brillouin sphere is the sphere centered around the origin of the body and still enveloping all of the body’s mass.
The latter two require the knowledge of the shape model of the target body, and their application comes with constraints like discretization \cite{wittick2017mascon} in the case of the mascon model or the assumption of homogeneous density in the case of polyhedral gravity models \cite{tsoulis2012analytical, tsoulis2021computational}.

With the recent advances in neural networks, a new approach was formulated, requiring no previous assumptions but only existing data originating from measurements or synthetic sources. With this data, one can train a neural network to solve the gravity inversion problem. Multiple approaches exist for creating such a network, as presented in \autoref{sec:related-work}. This paper focuses on geodesyNets —a network learning the density distribution of a body whose integration leads to gravity \cite{izzo2021geodesy}.
Previous work showed acceptable error boundaries for models trained with a synthetic ground truth based on a mascon model.
In addition, this approach was effectively used to explain the density of asteroid (101955) Bennu, with evidence from the OSIRIS-REx mission. However, the results were affected negatively due to noise in the input signal \cite{von2021study}.

Thus, this work aims to determine if this approach is still practical given that multiple sources of perturbations can pollute the gravity signal used for training, ranging from non-gravitational forces like solar radiation pressure to measurement inaccuracies. Further, the utilized gravity signal might be weaker due to safety distance when on a trajectory far from the celestial body, further hindering the network from learning correctly.
Last, it is investigated if pretraining on a prior shape model could improve the network performance under these constraints and if it reduces the amount of training iterations needed, thus reducing computational cost in a hypothetical onboard scenario.

This work uses a polyhedral gravity model and a mascon model for conducting these experiments. It demonstrates the applicability of geodesyNets trained with a noisy input signal, as long as the induced error is reasonably sized compared to the gravity signal. This means that an error relative to the magnitude of the gravity signal in the training data set still leads to acceptable errors. On the other hand, a large absolute limit on measurement accuracy can render any training useless.
This circumstance exemplifies in the conducted study about the sampling distance utilized for training the network. Finally, it is shown how pretraining can reduce the number of iterations required for training.

All results and code are made publicly available via GitHub.

\section{Related Work}
\label{sec:related-work}

\subsection{Polyhedral Gravity Models}

Polyhedral gravity models can calculate the full gravity tensor, including potential, acceleration, and second derivatives for an arbitrary given point $P$ around a polyhedral source. They provide an analytical solution given homogeneous density distributions and given a shape model.
One of the approaches to implementing a polyhedral gravity model is given by Tsoulis et al. \cite{tsoulis2012analytical, tsoulis2021computational}. They use a line integral approach to convert the triple integral into a nested summation, which, thanks to the introduction of dedicated singularity terms avoids potential singularities that can affect these models \cite{petrovic1996determination, tsoulis2001singularities}.
In this work, we use Tsoulis model to compare the performance of a geodesyNet trained with a mascon model to a network trained with a polyhedral model, but also to determine how the mesh granularity affects the achievable precision.
Furthermore, we perform a detailed study of the achievable precision at close range since the high accuracy, even within the Brillouin sphere, and the analytical formulation does not require numerical integration methods.

\subsection{Mascon Models}

Mascon models are the second approach for modeling gravity utilized in this work. Here, the body is represented as a combination of multiple point masses, so-called mascons (short for “mass concentrations” \cite{konopliv2001recent}) filling up its volume. The mascon elements do not need to be of uniform size. Instead, various approaches exist, combining mascons of different weights \cite{wittick2017mascon}. Its simplicity and ability to model irregular shapes and density distributions as they appear for small bodies make the model appealing. Regardless, high accuracy of the gravity field can only be achieved with an extensive number of mascons for non-spherical / irregular objects. Even then, the field’s accuracy near the body’s surface remains challenging due to the discretized mass distribution \cite{wittick2017mascon}.

\subsection{Gravity Modeling with Neural Networks}
\label{sec:related-work-other}

Artificial intelligence emerging in almost any area in recent years provides new opportunities in modeling traditionally expensive computational processes in physics \cite{izzo2022selected, xie2022neural}. One such domain is modeling the gravity of irregularly shaped bodies and mapping positions to accelerations.
In this context, one can distinguish between neural networks representing the actual body, enabling an indirect mapping to gravitational accelerations, and those directly mapping positions around a given body to gravitation.
Thus, the former can also be referred to as neural fields since they parameterize the bodies \cite{xie2022neural}.
A geodesyNet initially described by \cite{izzo2021geodesy} represents a body through its neural density field.

The inspiration for geodesyNets originates from Neural Radiance Fields (NeRF) introduced by \cite{mildenhall2021nerf}. Their network learns to represent a three-dimensional object from two-dimensional pictures. Thus, their network maps a 5D input vector for position and view angle to a 2D color and volume density vector. Similarly how they solve the inverse problem of image rendering, a geodesyNet solves the inverse problem of gravity inversion mapping a cartesian vector to a candidate body density \cite{izzo2021geodesy}. The neural field of a geodesyNet represents the body's density distribution.

The subsequent \autoref{sec:geodesynet} will describe it in detail since it is the object of study, whereas this section provides a brief look over alternatives from the first category of networks regressing acceleration on positions.

Generally, all of the following presented networks have in common that they directly map coordinates to potentials or accelerations and are trained on a polyhedral ground truth. However, the concrete approaches vary a lot. Furfaro et al. \cite{furfaro2021modeling} use Single-Layer Feedforward Networks using Extreme Learning, reducing the number of required iterations to fine-tune their model to map the relationship between spacecraft points and gravitational acceleration. They report relative errors (relRMSE) below $5\%$ for asteroid 25143 Itokawa and only slightly above $5\%$ for comet 67/P Churyumov-Gerasimenko. They conducted two experiments. One time the global gravity field was learned using target points sampled from a near-range sphere around the body. The second time a local field was learned using target points from a cylinder above a given landing zone.
Their approach contrasts this work as here the robustness is examined when also only sampling from far away.

Cheng et al. \cite{cheng2020real} use deep neural networks (DNN) with the aim of generating trajectories in a computationally cheap way for a soft landing on small bodies.
They show that the median error of their model is $0.33 \%$ when taking the mean over the three axial directions. Given that $1\%$ deviations are usually acceptable in traditional gravity models \cite{cheng2020real}, they conclude that their DNN is a practical approach. Similarly to the later employed geodesyNet, they normalize the input and outputs to the range $(-1, 1)$. Further, they do not train the network with sample points from the near field but instead with points some kilometers away from the body. In their case study for 433 Eros, points are collected from the sphere $3 \, km$ to $60 \, km$.

Finally, Gao et al. \cite{gao2019efficient} model the gravity field of multiple sample bodies with a  Gaussian Process Regression. They report a relative mean error of $1.27\%$ for their bodies when validating inside the sampling area. They only train their networks with samples including the near range of the bodies. However, in contrast to the study presented later on geodesyNet, their approach shows strong generalization difficulties when further away from the body. Thus, the relative error increases beyond $60\%$ in the case of 433 Eros at $40 \, km$ distance. For comparison: Training was done here with points up to $20 \, km$ distance.

Martin et al. \cite{martin2022earth, martin2022small, martin2023physics3} also present a continuously improved model that maps Cartesian coordinates to gravity. However, their approach differs from the usual training in the sense that the network is bound to additional constraints beyond the usual loss to guarantee physical correctness. These constraints are integrated into the loss function and include, for example, the satisfaction of Laplace's equation. The training thus penalizes not only inaccuracy but also physical violations of the underlying differential equations.
They also train on 433 Eros with sampling up to a distance of three times the radius. The network generalizes well to distance, but has difficulties predicting closer to the asteroid.

\section{Methods}
\label{sec:methods}

\subsection{Ground-Truth Models}

\autoref{eq:polyhedral-attraction} shows the derived formalism for computing the gravitational acceleration around a polyhedral source at the origin with $G$ as the gravitational constant, $\rho$ as the constant density using the polyhedral gravity model by Tsoulis et al. \cite{tsoulis2012analytical, tsoulis2021computational}.
It consists of two summations with the outer summing over the polyhedral faces $p$ and the inner one iterating over the segments $q$ of each face. For a detailed description, refer to Tsoulis et al. \cite{tsoulis2012analytical, tsoulis2021computational}.

\begin{equation}
	\overrightarrow{a} = G \rho \cdot \sum_{p = 1}^{n} \overrightarrow{N_p} \cdot \left[ \sum_{q = 1}^{m} \sigma_{pq} h_{pq} LN_{pq} + h_p \sum_{q = 1}^{m} \sigma_{pq} AN_{pq} + \sing_{\mathcal{A}_p} \right]
	\label{eq:polyhedral-attraction}
\end{equation}

This polyhedral gravity model was recently implemented with modern C++ in a work preceding this one \cite{schuhmacher2022efficient}.

This model will be contrasted with the mascon model, whose formula is displayed in \autoref{eq:mascon} for a generic point $\overrightarrow{r_i}$ and a set of mascons $\mathcal{M} = \{(x_i, y_i , z_i) \; i=1..n\}$, in the following experiments. The aim is to determine the achievable precision with the knowledge that the mascon model, used in the original paper, has difficulties in the immediate proximity to the surface due to discretization.
Next to the calculation of the ground truth is the evaluation of the model. For this purpose, the neuronal density field is integrated using \autoref{eq:numerical-integration}. Since the models were always trained with vectorial accelerations and not scalar potentials, demonstratively, only these formulas are displayed.

\begin{equation}
	\label{eq:mascon}
	\overrightarrow{a}(\overrightarrow{r_i}) = -G \sum_{j=1}^{n} \frac{m_j}{r^{3}_{ij}} r_{ij}
\end{equation}

\subsection{GeodesyNets - Neural Density Fields}
\label{sec:geodesynet}

A neural density field, also referred to as geodesyNet \cite{izzo2021geodesy, izzo2022selected}, is a fully-connected neural network trained with a gravity signal from an arbitrarily shaped body with the aim of learning the body's density distribution. Thus, it solves the gravity inversion problem and provides a fully differentiable expression for mapping a cartesian point to a candidate body density (see \autoref{eq:density}) compatible with the observed gravity signal.

\begin{equation}
	f(x,y,z) \rightarrow \rho
	\label{eq:density}
\end{equation}

The appeal of this method is that it does not necessarily require a shape model, as is the case with the aforementioned gravitational models, but that the gravitational signal can come from any source. Further, it converges even inside the Brillouin sphere and can learn heterogeneous density distributions.

Using the trained model, a numerical integration over the neural density field can be performed to calculate the potential or the acceleration. This procedure is displayed in \autoref{eq:numerical-integration}.

\begin{equation}
	\label{eq:numerical-integration}
	\overrightarrow{a} = G \int_{x \in V} \frac{\rho(x)}{|\overrightarrow{r} - \overrightarrow{x}|^{3}} (\overrightarrow{r} - \overrightarrow{x}) \; dV
\end{equation}

Previous studies utilized this training approach in combination with the mascon model as ground truth, demonstrating its theoretical applicability \cite{izzo2021geodesy}. The measured relative error was always less than $1\%$ even near bodies. There, however, it was not considered how robust the network would be if the sample points were at a distance of several kilometers as they have been in previous missions to minimize the risk of collision.

GeodesyNets have also been looked at from a practical point of view with measured gravity signals obtained from the OSIRIS-REx mission, demonstrating the practical applicability of the given approach while also giving first insights above limitations of the approach given the noise in a measured gravity signal \cite{von2021study}.

This work now provides a detailed study of the disruptive factors which could lead to wrong model behavior, including noise in the ground-truth measurements or errors in the numerical integration. As well as the influence that the distance has on the possible precision.

\begin{figure}[tbh!]
	\centering
	\includegraphics[width=0.75\linewidth]{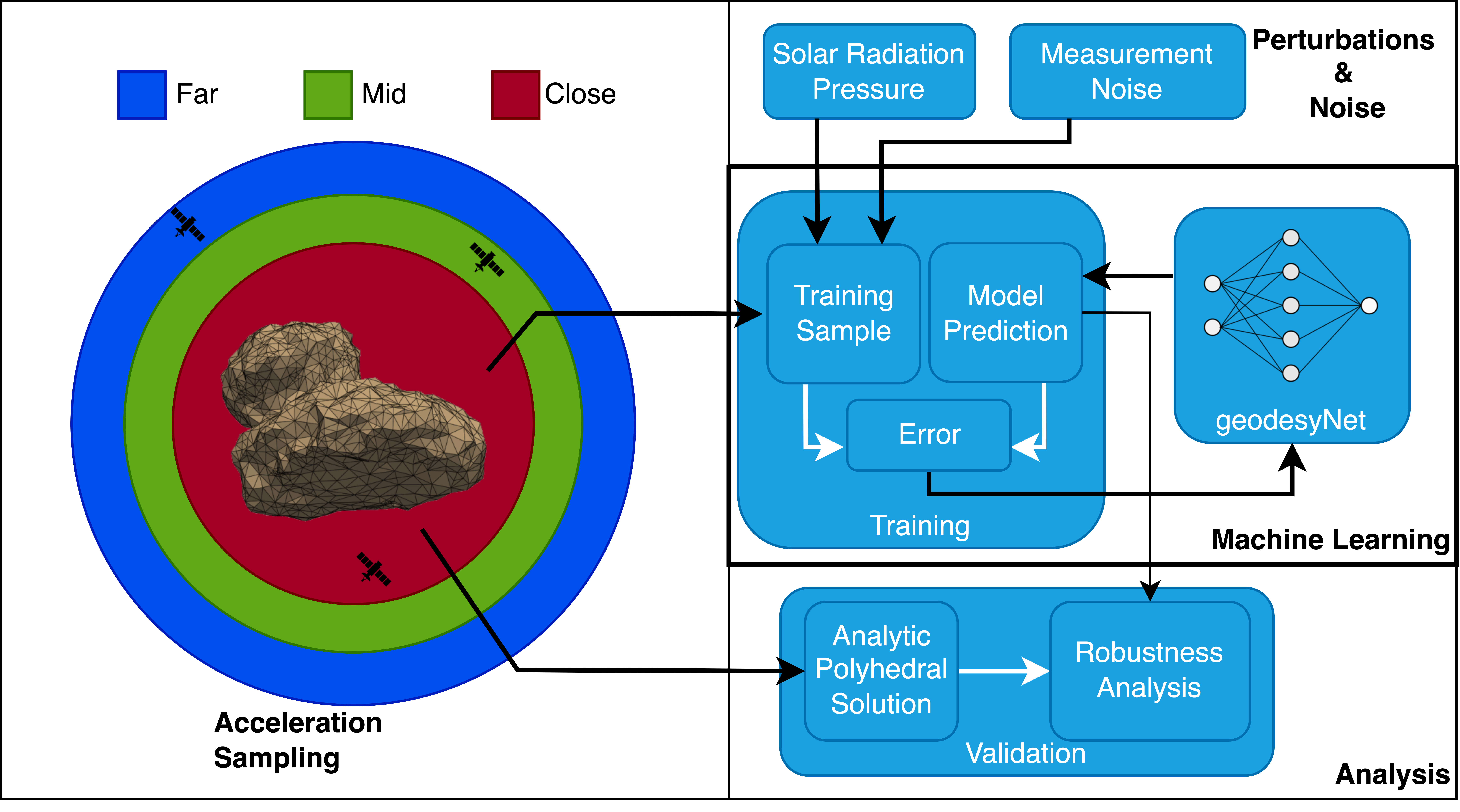}
	\caption{Training and validation toolchain: Sampling with noise and from different distances; Training for several thousand iterations; Comparing to the actual ground truth and calculation of relative error.}
	\label{fig:nn-training}
\end{figure}

\subsection{Sampling Points for Training}

To begin with, all meshes are normalized versions so that the bodies fit in the hypercube $[-1, 1]^3$
The spherical envelopes from which has been sampled for training were $(0,1)$, $(1,3)$, and $(3,5)$. The former covers the close range around the bodies, which the original study used for sampling. Here now, two more sampling distances are included. The mid-range $(1,3)$ and the far-range $(3,5)$ represent realistic distances an orbiter could reach around the body. The latter distances are the more realistic scenarios with respect to onboard training, as they maintain a certain safety distance for the spacecraft. In contrast, the close distance $(0,1)$ would only be practical in a gravity-model-based (pre-)training. In the following, it is refered to $(0,1)$, $(1,3)$, and $(3,5)$ as close-, mid- and far-range.

\begin{figure}[tbh!]
	\centering
	\begin{subfigure}{0.49\textwidth}
		\centering
		\includegraphics[width=\textwidth]{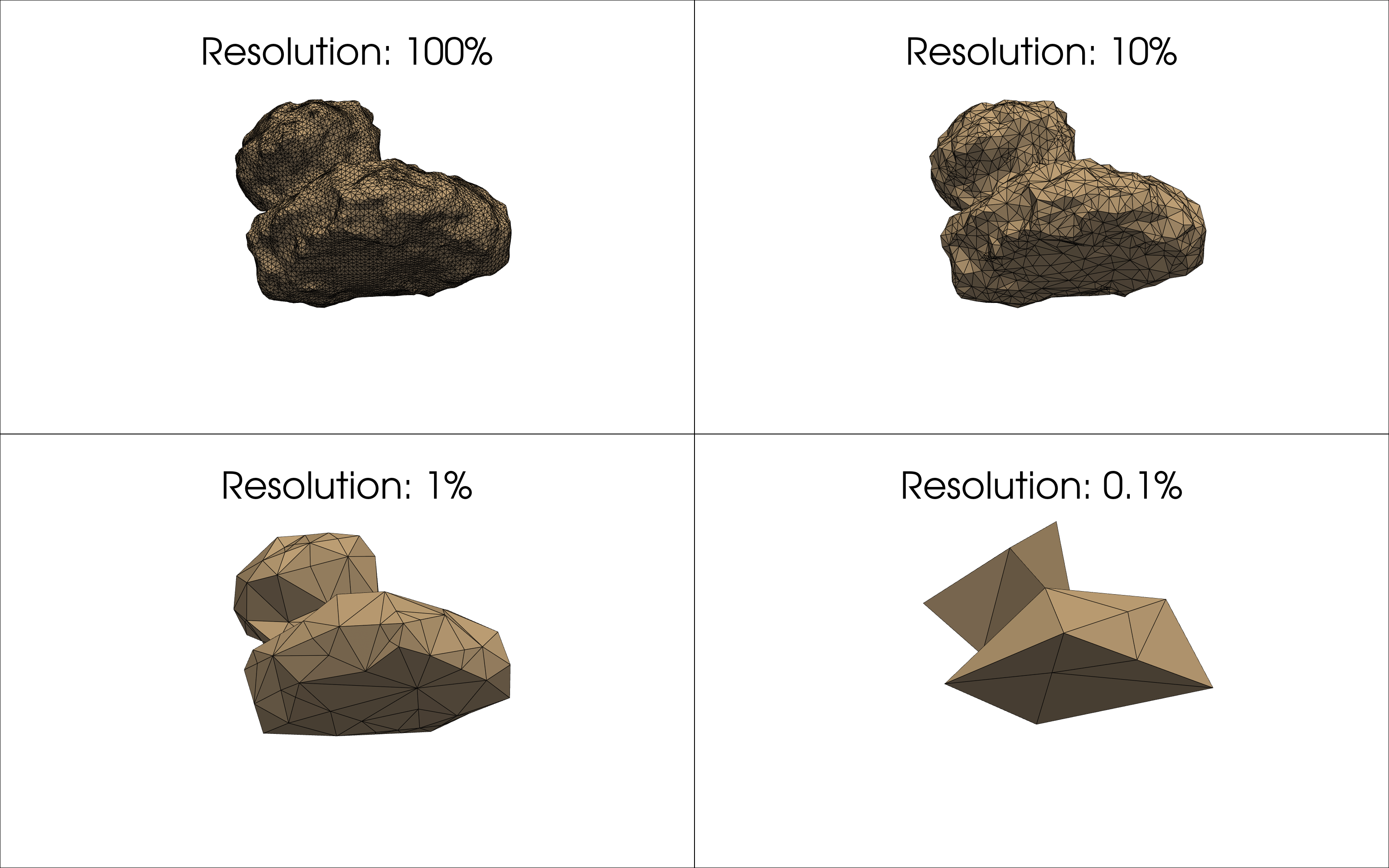}
		\caption{Churyumov-Gerasimenko}
	\end{subfigure}
	\hfill
	\begin{subfigure}{0.49\textwidth}
		\centering
		\includegraphics[width=\textwidth]{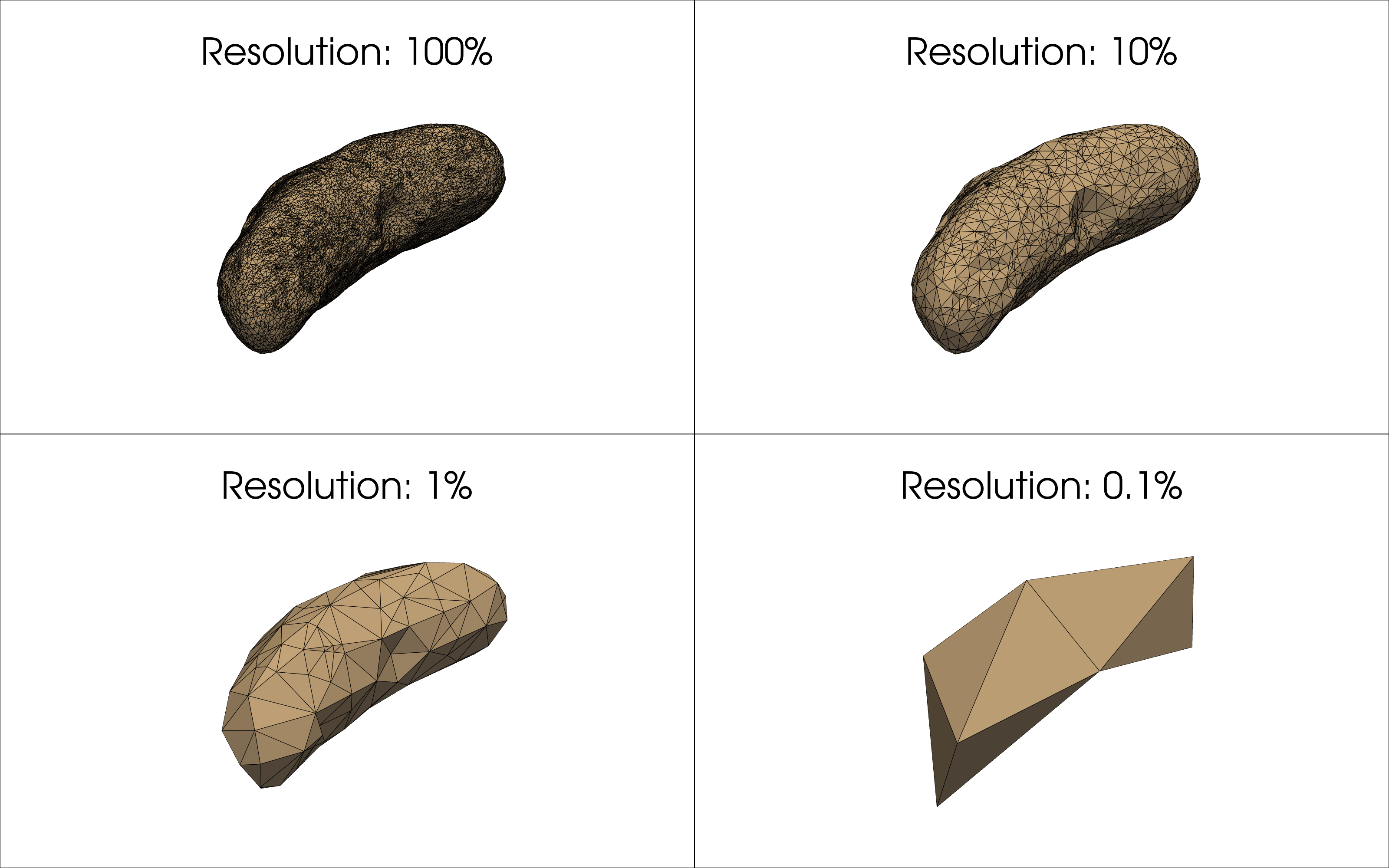}
		\caption{Eros}
	\end{subfigure}
	\hfill
	\caption{The original high resolution mesh denoted with $100\%$ and downscaled to meshes with respectively only $10\%$, $1\%$ or $0.1\%$ of vertices and faces.}
	\label{fig:body-mesh}
\end{figure}

The utilized mesh for Eros comes from Gaskell \cite{gaskell2008eros}, and the one for Churyumov-Gerasimenko from the European Space Agency  \cite{esa2019cgmodel}. Eros consists of 7374 vertices and 14744 triangular faces, whereas Churyumov-Gerasimenko comprises 9149 vertices and 18294 triangular faces. Both are based on the measurements of the probes that visited the bodies. These original meshes are now referred to as $100\%$ resolution meshes.
The mascon models used for both models are from Izzo \& Gómez \cite{izzo2021geodesy} and are derived from the $100\%$ resolution meshes. For this purpose, a centroid with mass $m_j$ is placed in each tetrahedron in the delaunay tetrahedralized version.
In the course of this study, additional meshes were constructed. These are downsampled versions with $10\%$, $1\%$, or $0.1\%$ of the vertices and faces. The purpose of this is to investigate to what extent a model can be (pre-)trained, even with a low resolution model as it may be available from astronomical observations. \autoref{fig:body-mesh} illustrates the employed input meshes for the polyhedral gravity model.

\subsection{Addition of Noise to the Ground Truth}

In addition to the ground-truth model, mesh resolution, and sampling distance, this study has one more parameter: noise. This characteristic is applied to the ground truth during the training. It thus represents that the measurements, the accelerations used for training, are subject to perturbations and measurement noise.

Three different types are investigated in the course of this study: a constant bias, additive Gaussian noise, and multiplicative Gaussian noise.

The constant bias represents the possible effects of solar radiation pressure (SRP) on the measurement results. In a previous study \cite{von2021study, leonard2019osiris}, this was a significant factor influencing the quality of the results. Although to subtract the SRP, it is necessary to know the mass and size of the object, along with its surroundings, such as the area facing towards the sun. Here, we will determine to what extent SRP influences the training. \autoref{eq:solar-pressure} \cite{kubo2023optimization} shows the employed approach for calculating the acceleration affecting the spacecraft with $P$ as the solar radiation pressure, $c$ as the speed of light and $G_{SC}$ as the solar constant, and $R$ as the distance from the sun in  astronomical units $[AU]$. \autoref{eq:srp-acc} calculates the acceleration $a$ given the area-to-mass ration of the spacecraft $A/m$.

\begin{align}
	\label{eq:solar-pressure}
	P &= \frac{G_{SC}}{c \cdot R^2} \\
	a &= P \cdot \frac{A}{m} \label{eq:srp-acc}
\end{align}

These equations are a simplified version of the calculation. Theoretically, the reflectivity, as given in Yousef et al. \cite{yousef2022balancing}, could also be included in the calculation. For this work, however, having only a baseline of the magnitude is sufficient. The calculated acceleration $a$ is then applied in one cartesian direction, assuming that the sun always shines from this direction.
In the following experiments, a value was used for the SRP as it could realistically affect a probe analogous to Rosetta. The mass was assumed to be $1422 \, kg$ (roughly Rosetta Dry Mass and Payload) and the area $69.88 \, m^2$. For the distance, the semi-major axis of the bodies was used in each case.
It should be added that the SRP for, e.g., a cubesat, would have a value of the same magnitude due to its smaller mass but equally smaller area.

Further, an additive Gaussian noise $X \sim \mathcal{N}(0, \sigma^2)$ is used during the training to simulate an absolute error in the measurement of the gravitational signal. The standard deviation $\sigma$ is chosen in a way that it simulates an absolute error of $10^{-4} \frac{m}{s^2}$ or $10^{-5} \frac{m}{s^2}$. After the meshes are normalized, this standard deviation has also been normalized.
The value $10^{-5}$ is the accuracy of GOCE (Gravity field and steady-state ocean circulation explorer) \cite{rummel2004high}. In addition to this value, $10^{-4}$ was chosen as the absolute precision to get an insight into how strongly one additional error magnitude affects the result accounting for even lower precision on spacecraft.

Further, a multiplicative Gaussian noise $X \sim \mathcal{N}(1, \sigma^2)$ is used during the training to simulate a relative error in the measurement of the gravitational signal. The assumption is that the magnitude of the gravitational signal is known, but the value can only be determined to a certain point. For example that the precision is limited relatively to the magnitude of the accelerations to $10^{-x}$ with $x \in \{1, 2, 3\}$.

\subsection{Training}

The toolchain used for training and performance evaluation is shown in \autoref{fig:nn-training}.  

This summarizes the characteristics of the training iterations. Overall, the same configuration is used for all other parameters as in the original paper \cite{izzo2021geodesy}. In particular, this includes the Mean Absolute Error (MAE) calculated from the ground truth $y$ and the model predictions $\hat{y}$ shown in \autoref{eq:loss}. Here, $\kappa$ (see \autoref{eq:kappa}) is a scaling parameter to normalize the mass and restrict training to learning the volume rather than finding the absolute mass \cite{izzo2021geodesy}.

\begin{align}
	\mathcal{L}_{\kappa \, MAE} &= \frac{1}{n} \sum_{i=1}^{n} | y_i - \kappa \hat{y_i} | \label{eq:loss} \\
	\text{with} \; \kappa &= \frac{\sum_{i=1}^{n} \hat{y_i} y_i}{\sum_{i=1}^{n} y_{i}^{2}} \label{eq:kappa}
\end{align}

\subsection{Validation}

For validation, the relative root mean squared error is used, which is calculated as shown in \autoref{eq:rel-rmse}. It is calculated for each model at the end of training for a range of distances from the body's surface. Comparisons are made against the polyhedral ground truth using the $100\%$ mesh resolution. The altitude sampler employed for this purpose uses the outward normal of the mesh faces to sample points at the appropriate altitude.
So the distance in the validation plot's x-axis is scaled by the altitudes from the surface, not the altitudes from the mathematical origin of the bodies.

\begin{equation}
	relRMSE = \sqrt{ \frac{1}{n} \sum_{i=1}^{n} \left( \frac{|y_i - \kappa \hat{y_i}|}{|| \hat{y_i} ||_{2}} \right)^{2} } \label{eq:rel-rmse}
\end{equation}

\section{Results}

For each scenario presented here, ten training runs have been performed with different seeds. The graphs show the mean value of the validation results and the standard deviation in a slightly transparent way.
The batch size has been set to $1000$ points, and every ten training iterations new point for training are sampled. For this purpose, a spherical point sampler randomly generates points in a spherical shell that do not lie within the sample body.

\subsection{Polyhedral vs. Mascon Model}
\label{sec:polyhedral-vs-mascon}

\begin{figure}[tbh!]
	\centering
	\includegraphics[width=0.9\textwidth]{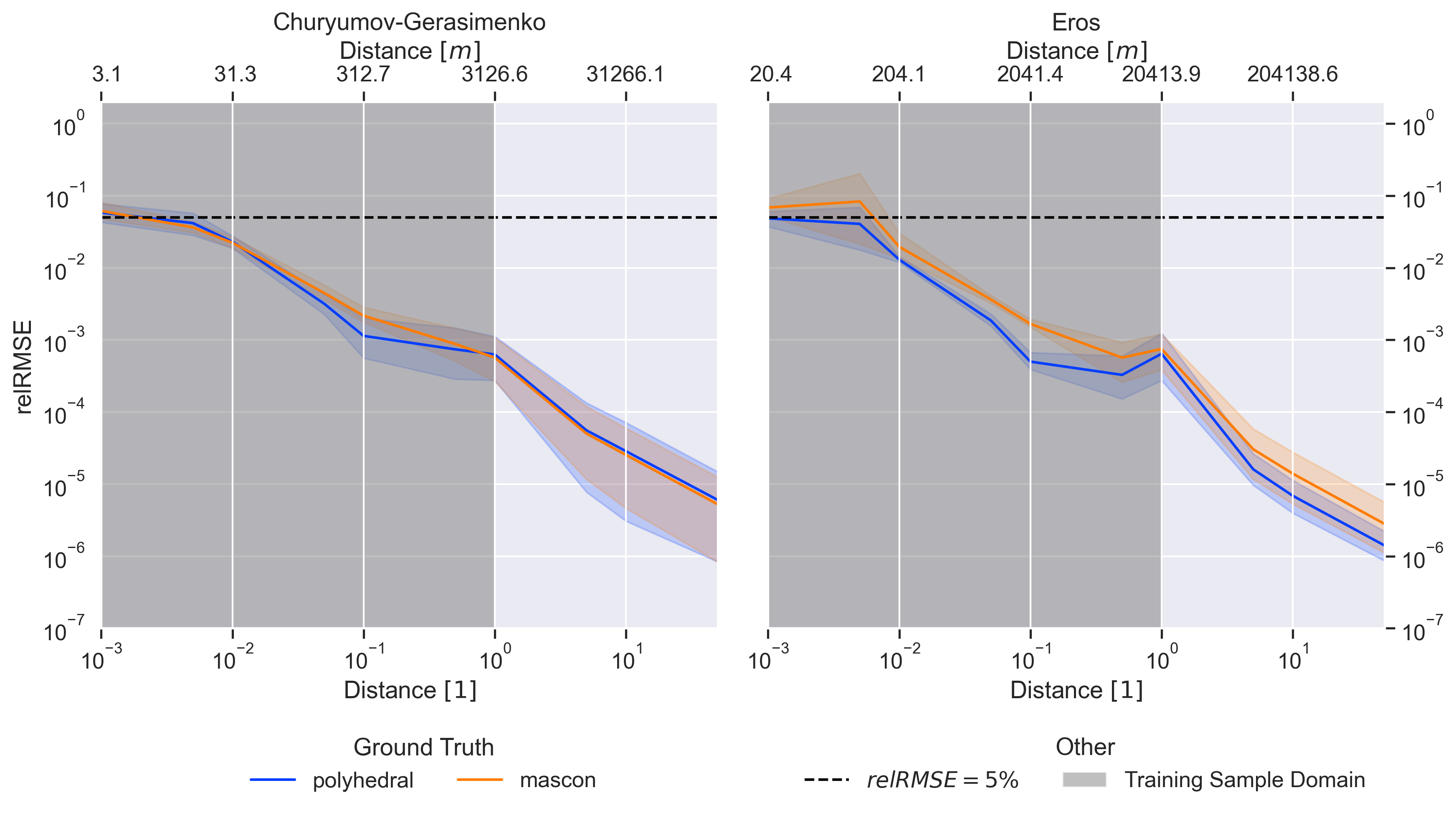}
	\caption{Relative errors of mascon-trained models and polyhedral-trained models on the $100\%$ resolution mesh}
	\label{fig:mascon-vs-polyhedral}
\end{figure}

\autoref{fig:mascon-vs-polyhedral} compares the relative errors of the models trained with a mascon model and a polyhedral model for the two studied bodies. Spherical sampling was performed in a shell $\in (0,1)$, but the results are similar for sampling with more distant shells. No qualitative difference can be found between models trained with the mascon or polyhedral models, not even in the close range around the body, as one would have expected due to the polyhedral model's improved accuracy in the close range.

For Churyumov-Gerasimenko, the relative errors are of similar size. E.g. for normalized distance $0.01$, the model's errors are close at $2.2\%$, whereas for distance $0.001$, the mascon models has a slightly larger mean error of $6.1\%$ compared to the polyhedral trained models' relative mean error of $5.8\%$.

For Eros, although the average relative error is slightly lower for the polyhedral model. The model's relRMSE trained with the polyhedral model is around $50\%$ to $70\%$ compared to the relRMSE of the mascon trained models through all validation distances.

\subsection{Sampling Distance}
\label{sec:sampling}

\begin{figure}[tbph]
	\centering
	\includegraphics[width=0.85\textwidth]{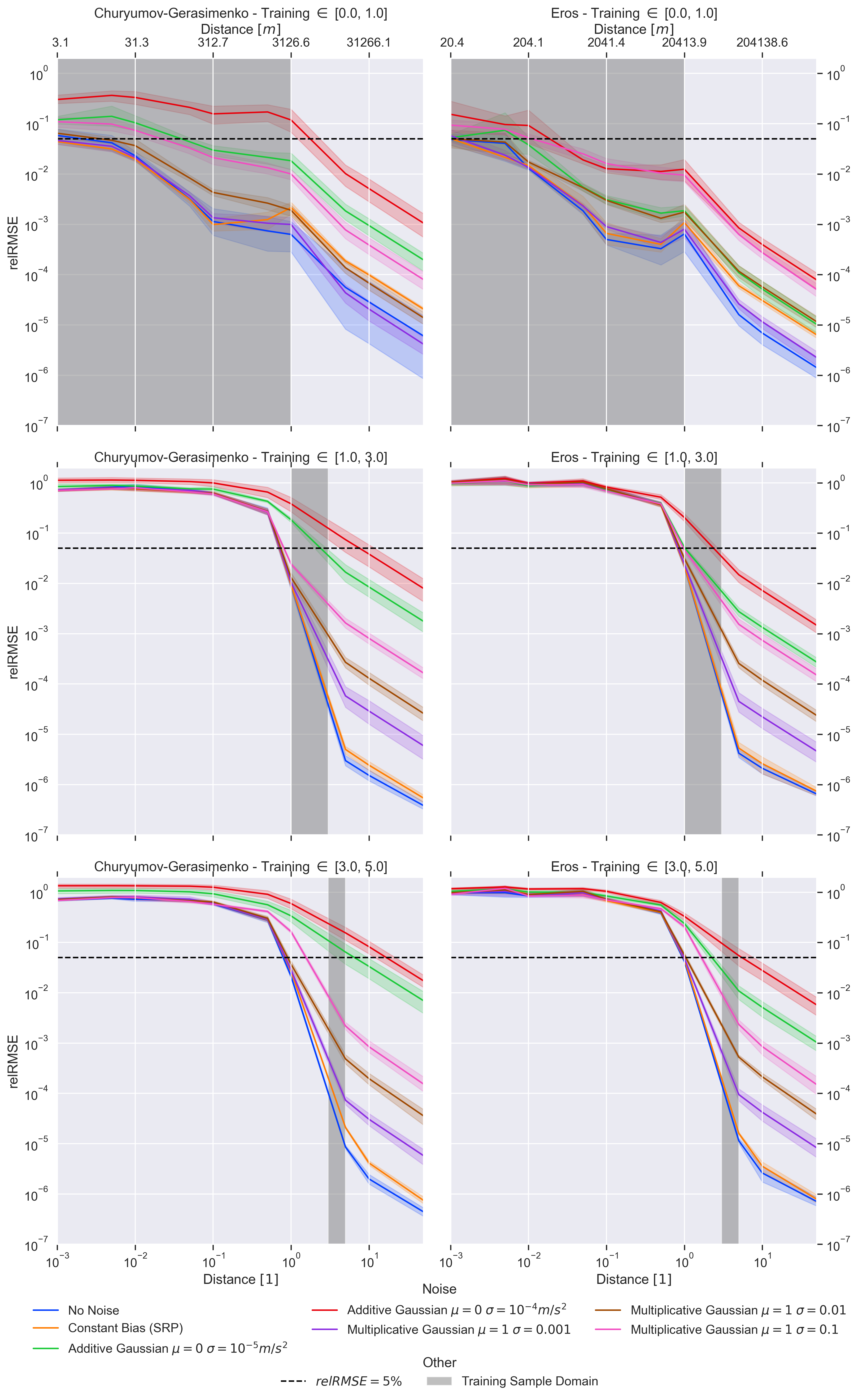}
	\caption{Effects of Noise and Sampling Shell on obtainable precision.}
	\label{fig:body-sampling-noise}
\end{figure}

Note that the blue lines in \autoref{fig:mascon-vs-polyhedral} and \autoref{fig:body-mesh-nn-performance} are the same as in \autoref{fig:body-sampling-noise} in the left column. They show the average relative error of ten trained models with the polyhedral ground truth, respectively, based on the $100\%$ resolution mesh and with sampling in the range $(0,1)$ during training.

In \autoref{fig:body-sampling-noise}, the lines show how the relative error changes for a given height above the target body when sampling only in the normalized range $(1,3)$ or $(3,5)$ during training.

Immediately, it is noticeable that if the network only gets points from mid or far away during the training, the relative error in the close range increases to $100\%$ and above. However, the network performs well for the mid to far range, and the mean relative error remains below $4.3\%$ until a normalized distance of $1$ or further away for models trained in mid or far range.
Hence, the network always performs well in the sampling range or further away in the observed cases. It generalizes to the ranges more distant than used during training.
It is also interesting to note that in the range where geometry does not play such a large role, the network is able to generalize and scale the predictions when getting closer to the body. This is reflected in the plots on the right side of \autoref{fig:body-sampling-noise}. The network has been trained with points from the region $(3,5)$, but the average relative error always remains below the $4.3\%$ mark, even for the $(1,3)$ region.

\subsection{Robustness to Noise}
\label{sec:noise}

This section refers again to \autoref{fig:body-sampling-noise}, which also displays the results when some noise has been applied to the ground truth for learning.

Starting with the constant bias, models trained with a constant bias approximately equal to the magnitude of the solar radiation pressure that would act on a probe such as Rosetta or a cubesat show an almost identical relative error in validation as models trained with the original ground truth.

In the case of additive Gaussian noise, which corresponds to an absolute error of $10^{-5} \, \frac{m}{s^2}$ or $10^{-4} \, \frac{m}{s^2}$ in non-normalized units, respectively, the network is most disturbed during learning. The mean relative errors are far beyond $5\%$ for an absolute error of $10^{-4} \, \frac{m}{s^2}$ even within the training range. However, the results differ for Churyumov-Gerasimenko and Eros. The network shows smaller relative errors for Eros than for Churyumov-Gerasimenko.

The multiplicative Gaussian noise that impacts the ground truth by $+/- 0.1\%, 1\%, 10\%$ respectively settles in the middle of the computed relative errors.
Training in the close range with a relative Gaussian noise of $1\%$ still results in relative errors below $4.7\%$ even until a range of $0.005$ for training in $(0,1)$. However, with $\sigma = 10\%$, the relative error maximally increases to $11.0\%$ for models trained in $(0,1)$.

\subsection{A posteriori: Using Pretraining}
\label{sec:pretraining}

After the previous sections have shown that noise and a long sampling distance increase the relative error, we will now investigate whether these results can be improved by pretraining.
The deliberation is to perform the pretraining on an imprecise low-resolution navigation model like it would be available due to observation before launch in the near range $(0,1)$. In the second step, the fine-tuning happens on the $100\%$ resolution mesh in the range $(3,5)$ for a limited number of iterations, effectively simulating a redefined improved mesh.
Even though we show results without noise terms here, experiments with those resulted in comparable results.

\begin{figure}[tbh!]
	\centering
	\includegraphics[width=0.9\textwidth]{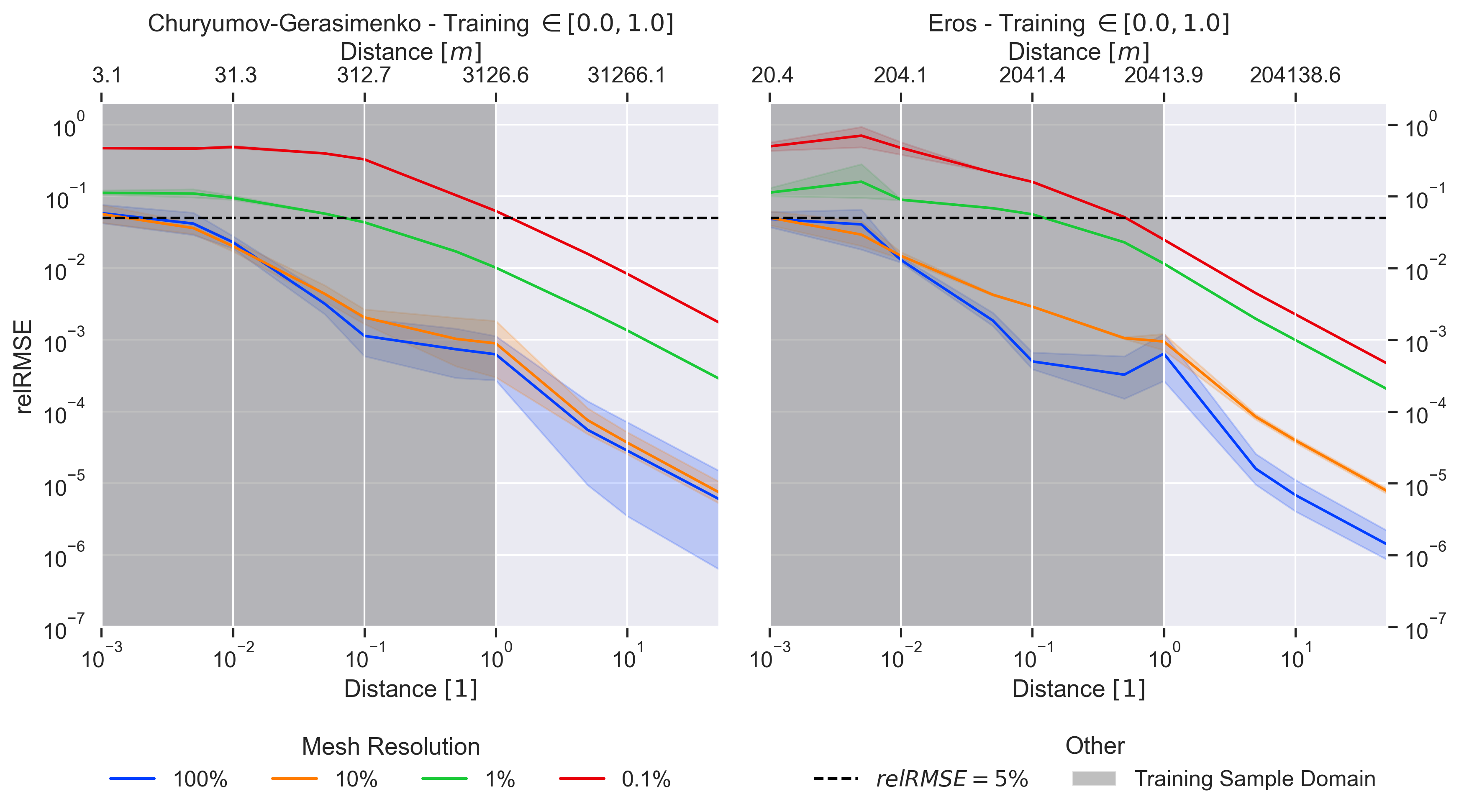}
	\caption{The performance of the $10 000$ iterations trained models depending on the utilized mesh resolution. The validation was done based on the $100\%$ polyhedral ground truth.}
	\label{fig:body-mesh-nn-performance}
\end{figure}

\autoref{fig:body-mesh-nn-performance} shows the achievable precision for the polyhedral model depending on the precision of the mesh used. A model trained with a mesh with only one-tenth of the vertices and faces achieves a comparable performance to the entire $100\%$ resolution mesh. Also interesting is that a geodesyNet trained with the $0.1\%$ resolution mesh still shows a solid relRMSE of only $6.3\%$ (Churyumov-Gerasimenko) and $2.5\%$ (Eros) for a normalized range of $1$.
The models displayed in \autoref{fig:body-mesh-nn-performance} are utilized as a pretraining base.

\begin{figure}[tbh!]
	\centering
	\includegraphics[width=0.9\textwidth]{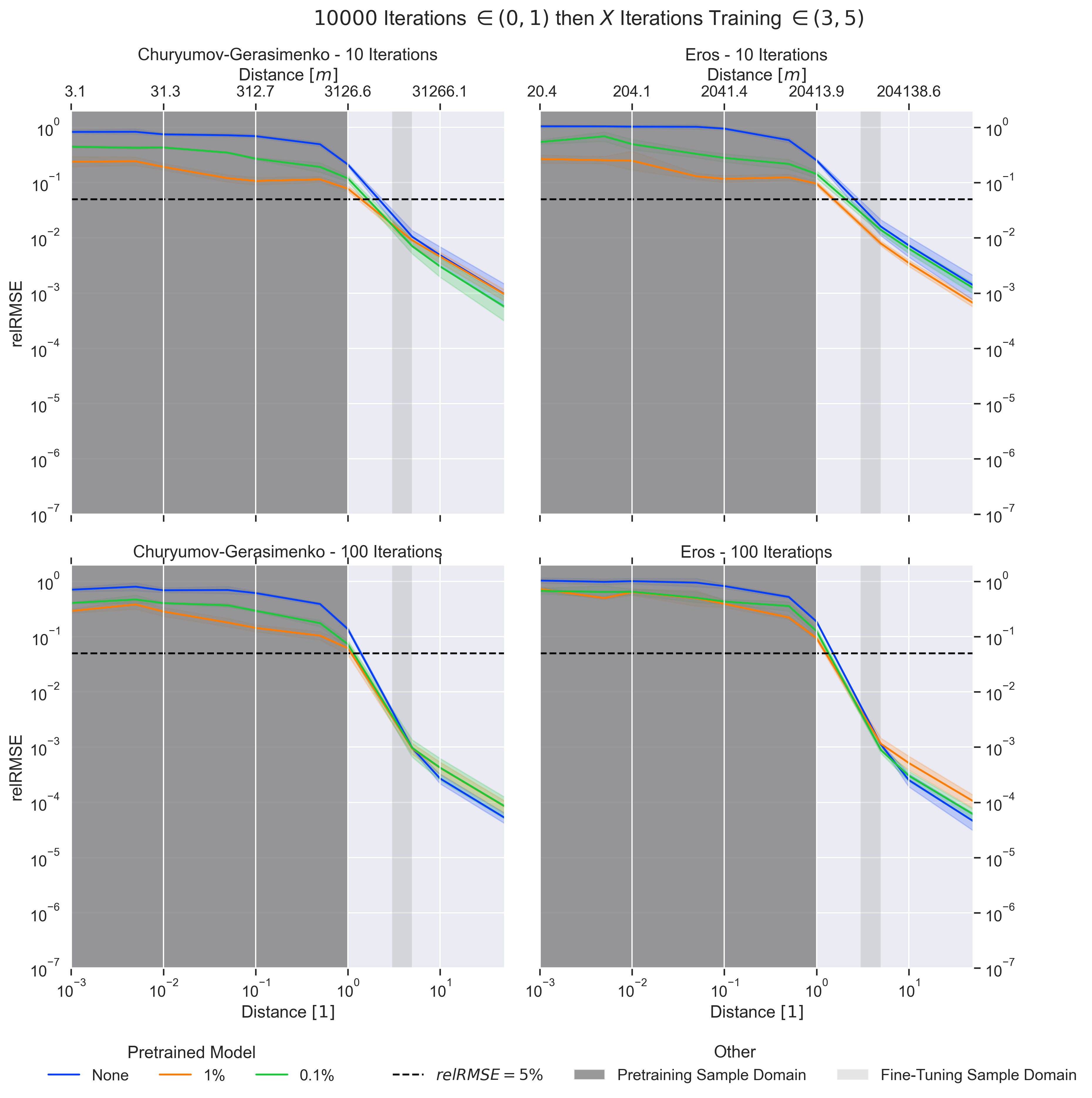}
	\caption{This figure compares the relative errors of pretrained models versus non-pretrained models. The pretraining was always conducted in the range $(0,1)$ marked in dark gray for $10 000$ iterations, and the fine-tuning (training for none) was conducted in the range $(3,5)$ marked in light gray for $10$ or $100$ iterations. Here, no noise was applied.}
	\label{fig:body-pretraining}
\end{figure}

\autoref{fig:body-pretraining} shows the results. Pretraining was performed for $10 000$ iterations, and fine-tuning (or training for the non-pretrained models) performed for $10$ or $100$ iterations. The pretrained models correspond to those in \autoref{fig:body-mesh-nn-performance}.

\autoref{fig:body-pretraining} shows that models pretrained on the near range data (green and orange curve)  overall outperform the models without any pretraining in the near range.
Further, the more precise the model utilized for pretraining, the better the obtainable precision in the near range.
As a second observation, it is noticeable that the relRMSE increases in the near range for the pretrained models if the fine tuning is conducted for 100 rather than 10 iterations (e.g. $0.1\%$ pretrained resolution, distance $0.1$: $28.1\% \rightarrow 43.2\%$) while the relRMSE decreases for the far field (e.g. $0.1\%$ pretrained resolution, distance $5.0$: $1.3\% \rightarrow 0.09\%$).
As the previous sections have shown, the model adapts to the far range when fine-tuning.

To summarize, pretraining allows better-performing models with fewer iterations. However, one has to be careful when fine-tuning with far-field data to maintain the generalization capabilities learned during pretraining in the near field.

\section{Discussion}

\subsection{Polyhedral vs. Mascon Model}

No clear qualitative difference could be found in the comparison between the mascon and polyhedral models in \autoref{sec:polyhedral-vs-mascon}. In both cases, the relative error is similar even for close distances. A possible reason could be that imprecision due to the number of vertices/edges in the mesh may be similarly limiting as the number of mascons. Furthermore, the numerical accuracy could be limited during training. Especially the numerical integration could be crucial and limit the maximum achievable precision. In addition, only single precision is used, but the polyhedral model provides a more precise result, i.e., precision that cannot be learned in this sense - since it cannot be represented in single precision. Mao et al. \cite{mao2020physics}, for example, could improve accuracy using double precision in their physics-informed neural network.

\subsection{Sampling Distance}

Regarding the results presented in \autoref{sec:sampling} on sampling, one could expect that the network could not predict the geometry and associated accelerations only by training in the mid-range to far-range. However, a geodesyNet is able to generalize and scale measurements as a function of distance successfully. This is clearly shown in the right plots in \autoref{fig:body-sampling-noise}. Conversely, a network is able to generalize successfully from the conducted sampling ranges up to $50$ times the distance. A possible reason for that could be the employed normalization and loss strategy involving the continuously re-calculated mass normalization factor.
Furthermore, the geodesyNet learns the shape of the body. Hence, it automatically satisfies the Laplace equation and thus represents a valid gravity field generated by the learned shape - a property enabling excellent generalization capability. In that sense, geodesyNets contrasts the in \autoref{sec:related-work-other} listed approaches using a regression.

The experiments also showed that the models could generalize from the far to the mid-range. As soon as the shape of the body strongly impacts the measured gravity signal, the generalization approaching closer to the body does not work anymore.
This feature could also be beneficial in a hypothetical onboard scenario, as it allows a probe to get closer step by step and improve the model accordingly. As here observed, the model makes a solid prediction a little closer than the actual sampling range during training.

\subsection{Noise}

\autoref{sec:noise} shows that a slight constant bias is calculable. In other words, future missions using a typical spacecraft will not have to worry about accounting for it.

It also shows that training a geodesyNet when expecting a sizeable absolute error in the measurements is a challenging task since one must find a way to overcome the noise. Otherwise, the convergence of the network to an acceptable solution is not achievable.
Especially if the error is notably large compared to the actual gravity; this is also why models trained for Eros perform better with additive Gaussian noise. Churyumov-Gerasimenko is less massive and smaller. Thus, the gravity signal is smaller than it would be in the case of Eros, and an absolute error dominates the input signal rather than merely perturbing it.

Suppose the magnitude of the gravity signal is known, and only the number of determinable digits is a problem, as in the case of multiplicative Gaussian noise in the $10\%$ deviation case. Even in that case, a robust network is constructible.

\subsection{Pretraining}
\label{sec:pretraining-disc}

\autoref{sec:pretraining} analyzed pretraining. 
A first result is that even with a ground truth of poor quality, as was the case here during training with the $0.1\%$ resolution mesh, a low error can be achieved up to the mid-range distance - an essential requirement for pretraining on Earth before the mission.

Further, the results show that pretraining reduces the model error and allows for obtaining a better performance with fewer gravity measurements. For example, \autoref{fig:body-pretraining} shows that the relRMSE in the case of Churyumov-Gerasimenko is with $11.9\%$ approximately half as large using pretraining with the $1\%$ resolution mesh than without any pretraining ($21.3\%$) in the same amount of iterations.

In an onboard scenario, where power consumption is of critical interest, pretraining on Earth, albeit with a lower resolution ground truth, is thus advantageous.

Moreover, it must be added in this context that the scenario, pretraining, and fine-tuning on range $(0,1)$ has not been considered here. Instead, the mixed range sampling scenario was considered, as it would be conceivable for a mission: pretraining before launch on a simple navigation model based on remote observation and slowly improving the model when approaching the target while re-training.

\section{Conclusion and Future Work}

In summary, this work studied the robustness of geodesyNets. The variables of interest were: the underlying training ground truth, the sampling distance for the training points, and the effects of different types of noise and whether pretraining can reduce the number of training iterations.

There were only minor differences when the geodesyNet was trained with the mascon or the polyhedral gravity model, probably due to limitations by the numerical precision and the relation of the mascon ground truth and the polyhedral mesh.

The geodesyNet cannot learn the geometry of an irregular body properly with only distant measurements.
However, the neural density field generalizes well and yields robust results even in areas where no training has been performed, as long as they are not in the close range. In order to be able to predict the acceleration in the close-range, near the surface, training in this region needs to be conducted.

Noise negatively impacts the training's results. However, this depends strongly on the magnitude compared to the input gravity signal. If there exists an absolute boundary for the measurable precision of the acceleration and thus the actual gravity signal is no longer distinguishable from the measurement error, the training is unproductive.
However, if the magnitude of the gravity signal is known, even with relative measurement deviations of up to $10\%$, solid training results are achievable.

From the point of view that the magnitude of the gravity signal is known, the training is successful.

Pretraining allows more precise results in an onboard scenario with less sampling and is preferable, even if the ground truth is of low resolution.

Future work could consider other forms of sampling, such as sampling with regard to realistic trajectories. Such an experiment requires a set of efficient trajectories maximizing the gravity signal - something currently being conducted in a related work by Mar{å}k et al. \cite{marak2023trajectory}.

\FloatBarrier

\printbibliography
\end{document}